\documentclass[11pt,pra,showpacs,showkeys]{revtex4}

\usepackage{amsfonts,amsmath,amssymb,amsthm}
\usepackage{bbm,bm}
\usepackage{graphics,graphicx}
\usepackage{latexsym}
\usepackage{color}
\usepackage{times}

\newcommand{\ra}{\rangle}
\newcommand{\la}{\langle}
\newcommand{\w}{\mathrm{W}}
\newcommand{\s}[1]{\sqrt{#1}}
\newcommand{\ket}[1]{\vert #1 \ra}

\newcommand{\ov}[2]{\left\la #1 | #2 \right\ra}
\renewcommand{\t}{\theta}
\renewcommand{\v}{\varphi}
\renewcommand{\o}{\otimes}

\begin{document}

\title{Universal behavior of the geometric entanglement measure of many-qubit  W states}
\author{Levon Tamaryan}
\affiliation{Physics Department, Yerevan State University, Yerevan, 375025, Armenia\\
Theory Department, Yerevan Physics Institute, Yerevan, 375036, Armenia}
\author{Zaruhi Ohanyan}
\affiliation{Department of Informatics and Applied Mathematics, Yerevan State University, Yerevan, 375025, Armenia}
\author{Sayatnova Tamaryan}
\affiliation{Theory Department, Yerevan Physics Institute, Yerevan, 375036, Armenia}
\begin{abstract}
We show that when $N\gg1$ the geometric entanglement measure of general N-qubit W states, except  maximally entangled W states,  is a one-variable function and depends only on the Bloch vector with the minimal $z$ component. Hence one can prepare a W state with the required maximal product overlap by altering the Bloch vector of a single qubit. Next we compute analytically the geometric measure of large-scale W states by describing these systems in terms of very few parameters. The final formula relates two quantities, namely the maximal product overlap and the Bloch vector, that can be easily estimated in experiments.
\end{abstract}

\pacs{03.67.Mn, 03.67.Bg, 05.70.Jk, 68.65.-k}
\keywords{entanglement manipulation, universality, multiparticle systems, geometric measure}

\maketitle

\section{Introduction}

The physics of many-particle systems differs fundamentally from the one of a few particles and gives rise to new interesting phenomena, such as phase transitions~\cite{crit-rev,orus} or quantum computing~\cite{shor-94,ek-91,tele-93,niels}. Entanglement theory, in particular, appears to have a much more complex and richer structure in the N-partite case than it has in the bipartite setting. This is reflected by the fact that multipartite entanglement is a very active field of research that has led to important insights into our understanding of many-particle physics~\cite{wei,ved,robust,cart,kryu,lip-una,shar,hub}. In view of this, it seems worthy to investigate also the behavior of entanglement measures for large-scale systems. Despite the fact that the number of entanglement parameters scales exponentially in the number of particles~\cite{sud}, it is sometimes possible to capture the most relevant physical properties by describing these systems in terms of very few parameters.

Recently a duality between highly entangled W states and product states has been established~\cite{dual}.
The important class of W states~\cite{w} represents a particular interesting set of quantum states associated with high robustness against particle loss and nonlocal properties of genuine entangled multipartite states~\cite{pop-w,sen,usha,par-08}. And different experimentally accessible schemes to generate multipartite W states have been proposed and put into practice over the years~\cite{wang,raz-02,jap,w-gen1}

The duality specifies a single-valued function $r$ of entanglement parameters. We shall refer to $r$ as the entanglement diameter, as it will play a crucial role throughout this article. Another reason for the term entanglement diameter is that $r$ can be interpreted geometrically as a diameter of a circumscribing sphere. The geometrical interpretation and its illustration will be presented in the appendix and now we focus on the physical significance of $r$.

The entanglement diameter uniquely defines the maximal product overlap and nearest product state~\cite{Shim,barn,bno,wei} of a given highly entangled W state. It has two exceptional points in the parameter space of W states. At the second exceptional point the reduced density operator of a some qubit is a constant multiple of the unit operator and then the entanglement diameter becomes infinite. The maximal product overlap $g$ of these states is a constant  regardless how many qubits are involved and what are the values of the remaining entanglement parameters. These states are known as shared quantum states and can be used as quantum channels for the perfect teleportation and dense coding. Thus the shared quantum states are uniquely defined as the states whose entanglement diameter is infinite.

Furthermore, highly entangled W states have two different entangled regions: the symmetric and asymmetric entangled regions. In the computational basis these regions can be defined as follows. If a W state is in the symmetric region, then the entanglement diameter is a fully symmetric function on the state parameters. Conversely, if a W state is in the asymmetric region, then there is a coefficient $c$ such that the $c$ dependence of the entanglement diameter differs dramatically from the dependencies of the remaining coefficients. Hence the point of intersection of the symmetric and asymmetric regions is the first exceptional point. It depends on state parameters and its role has not been revealed so far. One thing was clear that the first exceptional point does not play an important role for three- and four-qubit W states~\cite{tri,toward}.

In this article we show that the first exceptional point is important for large-scale W states. It approaches to a fixed point when number of qubits $N$ increases and becomes state-independent(up to $1/N$ corrections) when $N\gg1$. As a consequence the entanglement diameter, as well as the maximal product overlap, becomes state-independent too and therefore many-qubit W states have two state-independent exceptional points. The underlying concept is that states whose entanglement parameters differ widely, may nevertheless have the same maximal product overlap and this phenomenon should occur at two fixed points. This is an analog of the universality of dynamical systems at critical points. It is an intriguing fact that systems with quite different microscopic parameters may behave equivalently at criticality. Fortunately, the renormalization group  provides an explanation for the emergence of universality in critical systems~\cite{crit-rev,orus,orus-vid}.

The developed concept distinguishes three classes of W states. The first class consists of highly entangled W states which are below both exceptional points and then $r$ varies from $r_{\min}=1/2$ to $r_0\approx1/\s{3}+O(1/N)$. We will show that these states are in the symmetric region and their entanglement diameter is a slowly oscillating function on entanglement parameters. Accordingly, the maximal product overlap is an almost everywhere constant close to its greatest lower bound. Similar results have been obtained in Ref.\cite{lin}, where it is shown that almost all multipartite pure states with sufficiently large number of parties are nearly maximally entangled with respect to the geometric measure~\cite{wei} and relative entropy of entanglement~\cite{ved}. We will not analyze rigorously these states since they are too entangled to be useful in quantum information theory~\cite{agr}.

The second and most interesting class consists of highly entangled W states which are between two exceptional points and then $r$ varies from $r_0$ to infinity. These states are in the asymmetric region and the behavior of the entanglement diameter is curious. We will show that $r$ is a one-variable function in this case and depends only on the Bloch vector ${\bm b}$ of a single qubit. As a consequence $g$ depends only on the same Bloch vector too and its behavior is universal. That is, regardless how many many qubits are involved and what are the remaining $N-1$ entanglement parameters the function $g({\bm b})$ is common. We will compute analytically $g({\bm b})$ and thereby find the Groverian and geometric entanglement measures~\cite{bno,wei} for the large-scale W states even if neither the number of particles nor the most of state parameters are known.

The third class consists of slightly entangled W states which are above both exceptional points. In this case the maximal product overlap takes the value of the largest coefficient and these states do not posses an entanglement diameter. We will not analyze this trivial case, but will combine the functions $g({\bm b})$ for slightly entangled and highly entangled asymmetric W states and obtain an interpolating function $g({\bm b})$ valid for both cases. It is in a perfect agreement with numerical solutions and quantifies the many-qubit entanglement in high accuracy($\Delta g/g\sim10^{-3}$ at $N\sim10)$.

The importance of the interpolating formula in quantum information is threefold. First, it connects two quantities, namely the Bloch vector and maximal product overlap, that can be easily estimated in experiments~\cite{bloch,guh-06}. Second, it is an example of how do we compute entanglement of a quantum state with many unknowns. Third, if the Bloch vector varies within the allowable domain then maximal product overlap ranges from its lower to its upper bounds. Then one can prepare the W state with the given maximal product overlap, say $g_0$, bringing into the position the Bloch vector, say $g({\bm b_0})=g_0$.

This article is organized as follows. In Sec.II, we review the main results of Ref.\cite{dual}. In Sec.III, we consider two- and three-parameter W states in the symmetric region and show that all of these states are almost maximally entangled. In Sec.IV, we consider three- and four-parameter W states in the asymmetric region and compute explicitly their maximal product overlap. In Sec.V, we generalize the results of Sec.III and Sec.IV to arbitrary many-qubit W states. In Sec.VI, we discuss our results. In the appendix, we provide a geometrical interpretation for the entanglement diameter.

\section{Maximal product overlap of W states}

In the computational basis N-qubit W states can be written as
\begin{equation}\label{2.w}
\ket{\w_n}=c_1\ket{100...0} + c_2\ket{010...0} + \cdots + c_N\ket{00...01},
\end{equation}
where the labels within kets refer to qubits 1,2,...,N in that order. The phases of the coefficients $c_k$ can be absorbed in the definitions of the local states $\ket{1_i}(i=1,2,...,N)$ and without loss of generality we consider only the case of positive parameters. For the simplicity we assume that $c_N$ is the maximal coefficient, that is, $c_N=\max(c_1, c_2,\cdots,c_N)$.

The maximal product overlap $g(\psi)$ of a pure state $\ket\psi$ is given by
\begin{equation}\label{2.g}
g(\psi)=\max_{u_1,u_2,...,u_N}|\ov{\psi}{u_1u_2...u_N}|,
\end{equation}
where the maximization runs over all product states. The larger $g$ is, the less entangled is $\ket{\psi}$. Hence for a quantum multipartite system the geometric entanglement measure $E_g$ is defined as
$$E_g=-2\log g(\psi).$$

The maximal product overlap demarcates three different entangled regions in the parameter space of W states:
\begin{enumerate}
\item The symmetric region of highly entangled W states, where  $g(c_1,c_2,...,c_N)$ is a symmetric function on all coefficients $c_i$.
\item The asymmetric region of highly entangled W states, where the invariance of  $g(c_1,c_2,...,c_N)$ under the permutations of coefficients $c_i$ ceases to be true.
\item The region of slightly entangled W states, where the inequity $g^2(c_1,c_2,...,c_N)>1/2$ holds.
\end{enumerate}

The appearance of the three entangled regions is the consequence of the existence of the two critical values for the largest coefficient $c_N$.  The first critical value $r_1(c_1,c_2,...,c_{N-1})$ is the solution of
\begin{equation}\label{2.r1}
\s{r_1^2-c_1^2}+\s{r_1^2-c_2^2}+\cdots+\s{r_1^2-c_{N-1}^2}=(N-2)\,r_1,
\end{equation}
which always exists and is unique. Note that the first critical value $r_1$ for the coefficient $c_N$ depends on the remaining coefficients $c_i, i=1,2,...,N-1$ but does not depend on $c_N$.  Nonetheless we will use the abbreviation $r_1(c_N)\equiv r_1(c_1,c_2,...,c_{N-1})$ whenever no confusion occurs.

The second critical value $r_2(c_1,c_2,...,c_{N-1)}$ is given by
\begin{equation}\label{2.r2}
r_2^2=c_1^2+c_2^2+\cdots+c_{N-1}^2.
\end{equation}
In what follows we will use the abbreviation $r_2(c_N)\equiv r_2(c_1,c_2,...,c_{N-1)}$ for the simplicity.

The second critical value is always greater than the first one and thus there are three cases. The first case  is $c_N < r_1$ and the maximal product overlap is expressed via the fully symmetric entanglement diameter $r(c_1,c_2,...,c_N)$, which is the unique solution of
\begin{equation}\label{2.symr}
\s{r^2-c_1^2}+\s{r^2-c_2^2}+\cdots+\s{r^2-c_N^2}=(N-2)\,r.
\end{equation}
Then $g$ is given by
\begin{equation}\label{2.symg}
g^2=\frac{r^2}{2^{N-2}}\left(1+\s{1-\frac{c_1^2}{r^2}}\right) \left(1+\s{1-\frac{c_2^2}{r^2}}\right) \cdots \left(1+\s{1-\frac{c_N^2}{r^2}}\right)
\end{equation}
and is a bounded function satisfying the inequalities  $c_N^2 < g^2(c_1,c_2,...,c_N) < 1/2$.

The second case is $r_1<c_N<r_2$. In this case the entanglement diameter $r(c_1,c_2,...,c_N)$ is the unique solution of
\begin{equation}\label{2.asymr}
\s{r^2-c_1^2}+\s{r^2-c_2^2}+\cdots-\s{r^2-c_N^2}=(N-2)\,r
\end{equation}
where only the last radical has the $-$ sign. Then $g$ takes the form
\begin{equation}\label{2.asymg}
g^2=\frac{r^2}{2^{N-2}}\left(1+\s{1-\frac{c_1^2}{r^2}}\right) \left(1+\s{1-\frac{c_2^2}{r^2}}\right) \cdots \left(1-\s{1-\frac{c_N^2}{r^2}}\right),
\end{equation}
where again the negative root is taken from the last radical. The expression \eqref{2.asymg} also has an upper and lower bounds and the inequalities  $c_N^2<g^2(c_1,c_2,...,c_N)<1/2$ hold everywhere in the asymmetric region.

The third case is $c_N\geq r_2$ and $g$ takes the value of the largest coefficient in this case
\begin{equation}\label{2.slight}
g^2=c_N^2.
\end{equation}
Now $g$ is bounded below and satisfies the inequality $g^2>1/2$.

Despite the fact that there exist three different expressions for the maximal product overlap it is a  continuous function on state parameters. Indeed, at $c_N=r_1$ both Eqs. \eqref{2.symr} and \eqref{2.asymr} have the same solution $r=r_1=c_N$ and expressions \eqref{2.symg} and \eqref{2.asymg} for $g$ coincide. At $c_N\to r_2$ the solution of \eqref{2.asymr} goes to infinity, $r\to\infty$, and \eqref{2.asymg} asymptotically comes to \eqref{2.slight}. At this limit $g^2=c_N^2=r_2^2=1/2$ and thus the surface $g^2(c_1,c_2,...,c_N)=1/2$ separates out slightly and highly entangled W states.

\section{Symmetric entanglement region}
In this section we analyze the maximal product overlap of two- and three-parameter W states that belong to the symmetric region of entanglement and show that if all coefficients are small, then $r$ is a  slowly oscillating function close to $1/2$.

\subsection{Two parameter W states}

Equations \eqref{2.symr} and \eqref{2.asymr} are solvable for $N=3$ and the answer is~\cite{tri}
\begin{equation}\label{3.w3}
g=
\begin{cases}
 \enskip 2R, &{\rm if}\quad c_3^2\leq c_1^2+c_2^2\cr
 \enskip c_3, &{\rm if}\quad c_3^2\geq c_1^2+c_2^2
\end{cases}
\end{equation}
where $R$ is the circumradius of the triangle $c_1,c_2,c_3$.

When $N\geq4$ Eqs. \eqref{2.symr} and \eqref{2.asymr} cannot be explicitly solved to give analytic expressions for $r$ in terms of the coefficients $c_k$ unless the state posses a symmetry. For example, for $N=4$ the equations are solvable if any two coefficients coincide and unsolvable  if all coefficients are arbitrary~\cite{toward}.

However, when $N\gg 1$ the situation is different. In many cases one can derive approximate solutions that quantify the entanglement of W states in high accuracy. We will find such approximate solutions and compare them with the exact or numerical solutions.

Consider first a W states with $N=m+k$ qubits and coefficients
\begin{equation}\label{3.mk-c}
c_1=c_2=\cdots=c_m=a,\quad c_{m+1}=c_{m+2}=\cdots=c_{m+k}=b.
\end{equation}
When  $m>1$ and $n>1$ the state is in the symmetric region and Eq.\eqref{2.symr} is reduced to
\begin{equation}\label{3.mk}
m\s{r^2-a^2}+k\s{r^2-b^2}=(N-2)\,r.
\end{equation}
This equation is solvable by radicals. Setting $a=\cos\t/\s{m},\; b=\sin\t/\s{k}$ one obtains
\begin{equation}\label{3.rmk}
r^2=\frac{2Nmk-4(N-1)(m\cos^2\t+k\sin^2\t)+2mk(N-2)\s{D}}{16(N-1)(m-1)(k-1)},
\end{equation}
where
\begin{equation}\label{3.D}
D=1-\frac{N-1}{mk}\sin^22\t.
\end{equation}

At  $m=1$ or $k=1$ the denominator and numerator vanish in Eq.\eqref{3.rmk}, but their ratio gives the correct answer. We will not consider this case since it is analyzed in detail  in Ref.\cite{toward}.

\begin{figure}[ht!]
\begin{center}
\includegraphics[width=7.5cm]{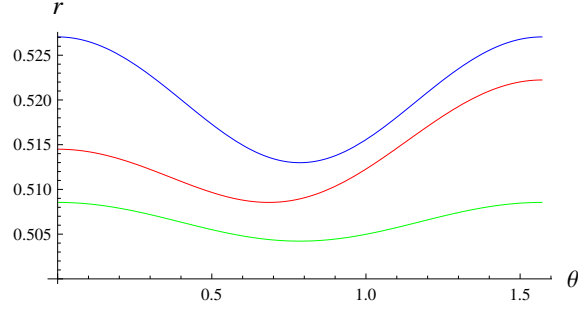}
\caption[fig1]{\label{r-mn}(Color online) The plots of the $\t$ dependence of the exact solution $r(\t)$ for the state \eqref{3.mk-c}. The top, middle and bottom lines represent the cases $(m=10,k=10),\;(m=12,k=18)$ and $(m=30,k=30)$, respectively.}
\end{center}
\end{figure}

If $m,k\gg1$, then $r$ is almost constant since
\begin{equation}\label{3.rapp}
r^2=\frac{1}{4}+O\left(\frac{1}{m}\right)+O\left(\frac{1}{k}\right).
\end{equation}
The question is when \eqref{3.rapp} achieves  the required accuracy. It can be understood by reference to Fig.\ref{r-mn}, where the $\t$ dependence of the exact solution \eqref{3.rmk} is plotted. The graphics show that $\Delta r/r\sim10^{-2}$ at $N\sim10$.
\begin{figure}[ht!]
\begin{center}
\includegraphics[width=7.5cm]{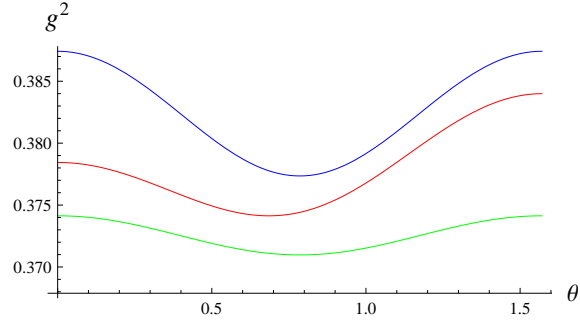}
\caption[fig2]{\label{g-mn}(Color online) The maximal product overlap function $g^2(\t)$ at different values of $m$ and $k$. The axes origin is put at the point $(0,1/e)$ to make it easer the comparison of the exact ant approximate solutions. The top, middle and bottom lines correspond to the values $(m=10,k=10),\;(m=12,k=18)$ and $(m=30,k=30)$, respectively.}
\end{center}
\end{figure}

As a consequence of Eq.\eqref{3.rapp} $g^2$ is also almost constant and close to its lower bound $1/e$~\cite{shim}. Indeed, using approximations
\begin{equation}\label{3.app-c}
  \frac{1}{2^m}\left(1+\s{1-\frac{a^2}{r^2}}\right)^m\approx e^{-ma^2/4r^2},\quad \frac{1}{2^k}\left(1+\s{1-\frac{b^2}{r^2}}\right)^k\approx e^{-kb^2/4r^2}
\end{equation}
one obtains
\begin{equation}\label{3.eapp}
g^2=\frac{1}{e}+O\left(\frac{1}{m}\right)+O\left(\frac{1}{k}\right).
\end{equation}

The behavior of the maximal product overlap $g(\t)$ given by Eqs. \eqref{2.symg} and \eqref{3.rmk} is plotted in Fig.\ref{g-mn}, which shows that $\Delta g/g\sim10^{-2}$ at $m,k\sim10$. It is difficult if not impossible to observe such small deviations of the maximal product overlap in experiments and therefore approximate formulas \eqref{3.rapp} and \eqref{3.eapp} have a good accuracy when $N\ge20$.

\subsection{Three parameter W states}

Consider now a three-parameter W state with $N=m+k+l$ qubits and coefficients
\begin{equation}\label{3.cmkl}
  c_1=\cdots=c_m = a,\;c_{m+1}=\cdots=c_{m+k} = b,\;c_{m+k+1}=\cdots=c_{m+k+l} = c.
\end{equation}

We will analyze the case $m,k,l\gg1$. Then Eq. \eqref{2.symr} can be rewritten as
\begin{equation}\label{3.mkl}
m\s{r^2-a^2}+k\s{r^2-b^2}+l\s{r^2-c^2}=(N-2)\,r.
\end{equation}
From the normalization condition $ma^2+kb^2+lc^2=1$ it follows that $a^2\leq1/m\ll1$ and similarly $b^2,c^2\ll1$. On the other hand \eqref{3.mkl} shows that $r\sim1$, and therefore we can expand the radicals in powers of $a^2/r^2,\;b^2/r^2$ and $c^2/r^2$. Then
\begin{equation}\label{3.rappr}
r^2=\frac{1}{4}+O\left(\frac{1}{m},\frac{1}{k},\frac{1}{l}\right).
\end{equation}
Again we got the same answer for $r$, which means that for partitions with large number of qubits $r$ depends neither on $m,k,l$ nor on $a,b,c$. More precisely, $r$ depends only on the expression $ma^2+kb^2+lc^2=|\psi|^2$, which drops out owing to the normalization condition.

\begin{figure}[ht!]
\begin{center}
\includegraphics[width=7.5cm]{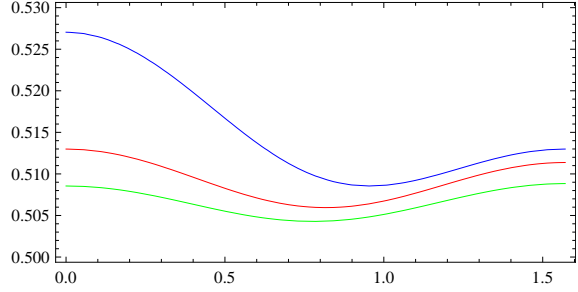}
\caption[fig3]{\label{r-mnl}(Color online)  The curves show the $\t$ dependence of the function $r(\t)$. The upper, middle and bottom curves represent the cases $(m=k=l=10,\v=\pi/4),\;(m=k=l=20,\v=5\pi/12)$ and $(m=10,k=20,l=30,\v=\pi/6)$, respectively.}
\end{center}
\end{figure}

The equation \eqref{3.mkl} can be solved explicitly, but the resulting half-page answer is impractical and we will compare \eqref{3.rappr} with the numerical solution instead. For this purpose we use the parametrization
$$a=\sin\t\cos\v/\s{m},\;b=\sin\t\sin\v/\s{k},\;c=\cos\t.$$
The behavior of the numerical solution $r(\t)$ of Eq.\eqref{3.mkl} for various values $m,k,l$ and $\v$ is plotted in Fig.\ref{r-mnl}. The graphics show that the approximate solution is in a perfect agreement with the numerical solution for $N\gg1$.

In summary, in the symmetric region of highly entangled W states the maximal product overlap does not depend on state parameters when many qubits are involved. Consider a W state, where $n_1,n_2,...,n_k$ product vectors in the computational basis have coefficients $c_1, c_2,...,c_k$, respectively. Then $g$ does not depend on partition numbers $n_i$ or amplitudes $c_i$  and the approximate solution \eqref{3.rapp} with the maximal product overlap \eqref{3.eapp} quantifies the entanglement in high accuracy. For example, at $N\sim 10$ the accuracy is $\Delta g/g\sim10^{-2}$. This approximation is true unless the condition $n_i\gg1(i=1,2,...,k)$ is violated. What is happening if this condition is violated, is analyzed in the next section.

\section{Asymmetric region of entanglement}
In this section we consider three- and four-parameter W states in the asymmetric region and show that if
one of coefficients exceeds the first critical value $r_1$, then $r$  is a rapidly increasing function and ranges from one-third to infinity when the maximal coefficient ranges from the first critical value to the second critical value.

\subsection{Three-parameter W states}

Consider now the case when $l=1$ in \eqref{3.cmkl}
\begin{equation}\label{3.cmk1}
  c_1=\cdots=c_m = a,\;c_{m+1}=\cdots=c_{m+k} = b,\;c_{m+k+1} = c.
\end{equation}


If $c\ll1$, then $c/r$ is small and $r$ is almost constant. This case is analyzed in the previous section and now we focus on the case when $c/r$ cannot be neglected. Then either $c\lesssim r_1$ or  $r_1<c<r_2$.

When $c\lesssim r_1$  Eq.\eqref{2.symr} takes the form
\begin{equation}\label{4.mk+1}
m\s{r^2-a^2}+k\s{r^2-b^2}+\s{r^2-c^2}=(N-2)\,r.
\end{equation}
The ratios $a/r$ and $b/r$ are small since $m,k\gg1$. Hence we expand the radicals in powers of these ratios up to  quadratic terms and solve the resulting equation. The answer is
\begin{equation}\label{4.r+1}
r=\frac{1}{2}\frac{1-c^2}{\s{1-2c^2}},\quad \s{1-\frac{c^2}{r^2}}=\frac{1-3c^2}{1-c^2},\quad \max(a^2,b^2)<c^2\leq\frac{1}{3}.
\end{equation}
It is reasonable that $r\to 1/2$ at $c\to0$.

When $c\geq r_1$  Eq.\eqref{2.asymr} takes the form
\begin{equation}\label{4.mk-1}
m\s{r^2-a^2}+k\s{r^2-b^2}-\s{r^2-c^2}=(N-2)\,r.
\end{equation}
Its approximate solution is
\begin{equation}\label{4.r-1}
r=\frac{1}{2}\frac{1-c^2}{\s{1-2c^2}},\quad \s{1-\frac{c^2}{r^2}}=\frac{3c^2-1}{1-c^2},\quad \frac{1}{3}<c^2<\frac{1}{2}.
\end{equation}
As one would expect, $r\to\infty$ at $c^2\to1/2$.

\begin{figure}[ht!]
\begin{center}
\includegraphics[width=7.5cm]{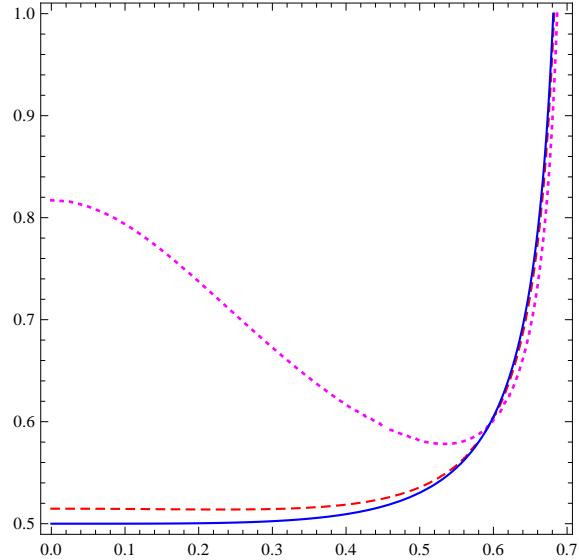}
\caption[fig4]{\label{r-mn1}(Color online)  Graphic illustrations of the function $r(c)$ for the three- and four-parameter W states. The solid curve is the approximate solution \eqref{4.r}. The dashed curve is the joined numerical solution of Eqs. \eqref{4.mk+1} and  \eqref{4.mk-1}. All remaining coefficients are 	 well away from the first critical value $(\approx0.58)$ when $c$ varies within the range of definition in this case. Accordingly, the state is in the symmetric region when $0<c<0.58$ and in the asymmetric region when $0.58<c<0.707$. The dotted line is the numerical solution for the state \eqref{3.cm+1-1}. Now another coefficient may exceed the first critical value. Therefore there are two first critical values, for the last and the preceding coefficients, respectively. The first critical value for the next to last coefficient $c$ is $\approx0.606$ and for the last coefficient $d$ is $\approx0.59$ which is attained at $c=0.45657$.  Thus the state is in the symmetric region when $0.45657<c<0.606$ and in the asymmetric region otherwise. Remarkably, the three curves coincide when $c>0.606$.}
\end{center}
\end{figure}

Surprisingly,  both solutions \eqref{4.r+1} and \eqref{4.r-1} can be unified to a single solution as follows
\begin{equation}\label{4.r}
r=\frac{1}{2}\frac{1-c^2}{\s{1-2c^2}},\quad \max(a^2,b^2)<c^2<\frac{1}{2}.
\end{equation}
The question at issue is when \eqref{4.r} gives a required accuracy in the asymmetric region $r_1<c<r_2$. We compare it with the numerical solutions of \eqref{4.mk+1} and \eqref{4.mk-1} for the values $(m=8,k=10,a/b=0.8,r_1^2\approx0.34)$ in Fig.\ref{r-mn1}, where the solid line is the plot of \eqref{4.r} and the dashed line is the numerical solution. Remarkably, the approximate solution is in a perfect agreement with the numerical one in the asymmetric region.

\subsection{Four-parameter W states}
However, there are W state that are outside the realm of the model sketched in the previous subsection. These are states with few (at most three) coefficients close to the first critical value $r_1\sim1/\s{3}$. In this case these coefficients are not small and the resulting $r$ should has a different behavior.

Notice, two coefficients cannot exceed the first critical value simultaneously. But we can construct W states whose coefficients depend on a free parameter in such a way that at one value of the free parameter the last coefficient exceeds the first critical value and at another value of the free parameter the preceding coefficient exceeds the first critical value. Below we construct an illustrative example of a such state and analyze its entanglement diameter.

An example is the 19-qubit four-parameter W state with coefficients
\begin{equation}\label{3.cm+1-1}
 c_1=\cdots=c_7\equiv a,\;c_{8}=\cdots=c_{17}\equiv b,\; c_{18}\equiv c,\;c_{19}\equiv d.
\end{equation}
For the normalized states we can use free parameters $\v,\;k$ and $c$  as follows
$$a^2 = \frac{\cos^2\v}{7k}(1-c^2),\; b^2 = \frac{\sin^2\v}{10k}(1-c^2),\;d^2 = \frac{k-1}{k}(1-c^2).$$

Now we analyze the function $r(c)$ at $k=1.8, \v=\pi/4$.

\begin{enumerate}
\item The  next to last coefficient $c$ coincides with its first critical value $r_1(c)$  at $c\approx0.606$ , that is, the solution of the system $$7\s{r_1^2-a^2}+10\s{r_1^2-b^2}+\s{r_1^2-d^2}=17r_1\;\, {\rm and}\;\, r_1=c$$ is $r_1=c\approx 0.606$. Then $r(c)$ should range from $r_1(c)$ to infinity when $c$ ranges from $r_1(c)$ to 1/2 and should has a vertical asymptote at $c^2\to1/2$.
\item The last coefficient $d$ coincides with its first critical value $r_1(d)$ at $d\approx0.593$, that is, the solution of the system
     $$7\s{r_1^2-a^2}+10\s{r_1^2-b^2}+\s{r_1^2-c^2}=17r_1\;\, {\rm and}\;\, r_1=d$$  is $r_1 = d \approx 0.593$. Note that at this point $c\approx0.45657$. Then $r$ should increase when $d$ ranges from $r_1(d)$ to $d_{\max}$. But the maximum value of $d$ is less than the second critical value since  $d_{\max}^2=d^2(c=0)=(k-1)/k=4/9<1/2$. Therefore $r$ should be bounded above in the interval $[r_1(d),d_{max}]$ and attain a maximum at $d_{max}$. As $d$ is a decreasing function on $c$, $r$ should attain a maximum at $c=0$ and then decrease when $c$ ranges from 0 to 0.45657.
\item The state is in the symmetric region when $d<r_1(d)$ and $c<r_1(c)$. Hence $r(c)$ should be minimal and nearly constant when $0.45657<c<0.606$.
\end{enumerate}

The dotted line in Fig.\ref{r-mn1} represents the $c$ dependence of the function $r(c)$. It agrees completely with the above analyze.

The main point is that all the three curves coincide when $c>r_1(c)$. In the next section we will show that this is not accidental and the curves must coincide. In this context the equation \eqref{4.r} is a surprising result. The quantity $r$, as well as the maximal product overlap $g$, depends from $c$ only. The rest of the state parameters appear in \eqref{4.r} in the combination $|\psi|^2-c^2$ and drop out by the normalization condition!
\begin{figure}[ht!]
\begin{center}
\includegraphics[width=7cm]{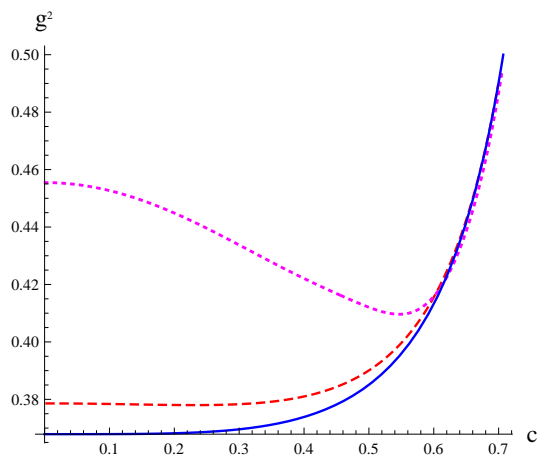}
\caption[fig5]{\label{g-mn1}(Color online) The plots of the function $g(c)$. The solid line is the approximate solution \eqref{4.g-as-ap}, the dashed and dotted lines are the numerical solutions for the states \eqref{3.cmk1} and \eqref{3.cm+1-1}, respectively. The curves may have different behaviors when $c_N<r_1$, but coincide when $c_N\geq r_1$.}
\end{center}
\end{figure}

Furthermore, we can derive an analytic expression for the maximal overlap. Using approximations \eqref{3.app-c} one obtains
\begin{equation}\label{4.g-as-ap}
g^2(c)=(1-c^2)e^{-(1-2c^2)/(1-c^2)}.
\end{equation}

The behavior of the function $g(c)$ is shown in Fig.\ref{g-mn1}. The solid line is the curve \eqref{4.g-as-ap}, the dashed curve is the numerical solution for the state \eqref{3.cmk1} and the dotted line is the numerical computation for the state \eqref{3.cm+1-1}. They all coincide when $c>r_1(c)$.

For highly entangled states the maximal product overlap ranges from its lower to the upper bound  when $c$ ranges from $r_1$ to $r_2$. On the other hand the Bloch vector ${\bm b}$ of $N$th qubit is collinear with axis $z$ and $b_z=1-2c^2$.  Thus $g$ is a one-variable function on $b_z$ and one can vary the entanglement of the multiqubit W state by altering the Bloch vector of a single qubit. The remaining qubits should be present in order to create an entanglement, but their individual characteristics do not play any role within the domain $-1<b_z<1-2r_1^2,N\gg1$. These qubits are just spectators, they should appear in the W state, but have no influence on the entanglement of the state.

\section{General case}

The results of the previous sections are based on the fact that the entanglement diameter $r$ is bounded below. In the symmetric region it is rigidly bound by the following theorem.

{\bf Theorem 1.} If $r$ is a solution of Eq.\eqref{2.symr}, then
\begin{equation}\label{4.bounds}
\frac{1}{4}\leq r^2\leq \frac{1}{2}.
\end{equation}

{\bf Proof.} Note that
$$\frac{c_i^2}{r^2}=\left(1+\s{1-\frac{c_i^2}{r^2}}\right)\left(1-\s{1-\frac{c_i^2}{r^2}}\right) \le 2\left(1-\s{1-\frac{c_i^2}{r^2}}\right).$$
By summing over $i$ the above inequality and using \eqref{2.symr} and the normalization condition one obtains
$$\frac{1}{r^2}\leq 2(n-n+2)=4.$$
Hence $r^2\geq1/4$. Next, from $x \le \sqrt{x}$ for $0 \le x \le 1$ it follows that
$$\sum_{i=1}^n\left(1-\frac{c_i^2}{r^2}\right)\le\sum_{i=1}^n\s{1-\frac{c_i^2}{r^2}},\quad
{\rm or}\quad
n-\frac{1}{r^2}\leq n-2,$$
that is, $r^2<1/2$.

The inequalities \eqref{4.bounds} allow us to understand the behavior of $g$ of arbitrary N-qubit W states in the symmetric region. Indeed, in this region $c_i^2\sim1/N$ and therefore $c_i^2/r^2\ll1$. Then one can expand the radicals in  \eqref{2.symr} and obtain
$$N-\frac{1}{2r^2}\approx N-2,$$
which generalizes \eqref{2.symr} and \eqref{2.symg} to arbitrary W states with $c_N\ll1$.

In Eq.\eqref{3.cmkl} we have chosen equal coefficients in order to reduce the number of independent parameters and make it easier the analyze. Now Theorem\;1 states that it is irrelevant whether some coefficients coincide. Decisive factor is that the coefficients $c_i$ are small($\sim1/\s{N}$). Then the ratios $c_i/r$ are small since $r$ is bounded below $(\sim1/2)$ and we can keep first nonvanishing orders of these ratios. Surprisingly, all these ratios are combined in such a way that they yield the Euclidean norm of the state function and the final answer becomes independent on the state parameters as well as the number of particles involved.

In the asymmetric region the entanglement diameter $r$ should has a lower bound but has not an upper bound since $r\to\infty$ at $c_2\to r_2$. One may expect that the lower bound of $r$ in the asymmetric region coincides with the upper bound of $r$ in the symmetric region. But the following theorem shows that this is not the case.

{\bf Theorem 2.} If $r$ is a solution of Eq.\eqref{2.asymr}, then
\begin{equation}\label{bounda}
r^2\geq \frac{1}{3}.
\end{equation}

{\bf Proof.} We use the same technique, namely
$$\frac{1}{r^2} = \sum_i^{N-1}\frac{c_i^2}{r^2} + \frac{c_N^2}{r^2} \le \sum_i^{N-1}2\left(1-\s{1-\frac{c_i^2}{r^2}}\right) + \frac{c_N^2}{r^2},$$
or
$$\frac{1}{r^2} \le 2-2\s{1-\frac{c_N^2}{r^2}} + \frac{c_N^2}{r^2} \le 3\quad {\rm since}\quad c_N\le r.$$
This bound, as well as bounds \eqref{4.bounds}, is tight, for example, $r^2\to1/3$ at $c^2\to1/3$ in \eqref{4.r}.

Theorem\,2 explains why the asymmetric approximate solution \eqref{4.r} fits the numerical date more quickly ($N\sim10$) than the symmetric one \eqref{3.rapp}($N\sim20$). First, the lower bound of $r$ is greater in this case. Second, since $c_N$ is greater($c_N>r_1$) the remaining coefficients should be smaller due to the normalization condition. These two factors together make the ratio $c_i/r$ smaller. Hence the approximate solution should has a better agreement with the exact one. Aside from that, $r$ is a fast increasing function and goes to the infinity unlike to the symmetric case. Hence the values of the  coefficients $c_i$ become irrelevant when $r\gg1$.

In fact  there is no W state in the asymmetric region that differs markedly from the above model when many qubits are involved. The following theorem completes the proof that in the asymmetric region the maximal product overlap is a one-variable function.

{\bf Theorem 3.} If $c_N=r_1$, then
\begin{equation}\label{bound-cr1}
r_1^2=\frac{1}{3}+O(\frac{1}{N})
\end{equation}

{\bf Proof.} Note that on the boundary of the symmetric and asymmetric regions $r=r_1=c_N$ and therefore $r_1^2\geq1/3$. Expanding the radicals in \eqref{2.r1} in powers of $c_i^2/r_1^2$ one obtains
$$N-1-\frac{1-c_N^2}{2c_N^2}+O(\frac{1}{N})=N-2,$$
which gives \eqref{bound-cr1}.

Now we are ready to explain what is happening in the asymmetric region.
\begin{enumerate}
\item When many qubits $(N\gg1)$ are involved the first critical value depends neither the number of qubits nor the state parameters and is a constant, $r_1\approx1/\s{3}$.
\item Regardless what is happening in the interval $0<c_N<r_1$ all functions $r(c)$ must  converge to the point $r(1/\s{3})\approx1/\s{3}$. This is the effect of the first critical value.
\item All functions $r(c)$ have the the same vertical asymptote, namely, $r(c)\to\infty$ at $c\to1/\s{2}$. This is the effect of the second critical value.
\end{enumerate}

\begin{figure}[ht!]
\begin{center}
\includegraphics[width=8cm]{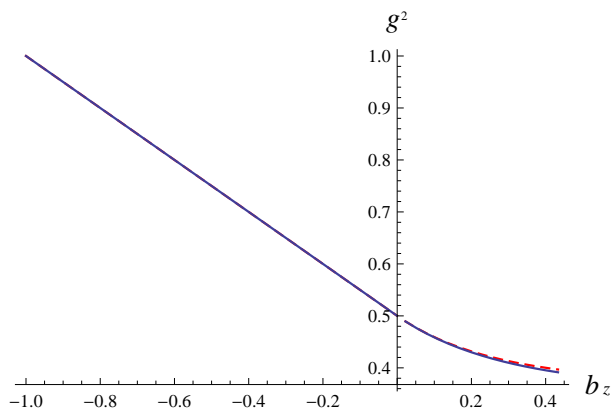}
\caption[fig6]{\label{int}(Color online) The maximal product overlap $g$ as a function of $z$ component of the Bloch vector $b_z$. The solid line is the interpolating formula \eqref{5.g}. The dashed line is the numerical computation for a 10-qubit W state.}
\end{center}
\end{figure}

These statements together give no chance to differ markedly exact and approximate solutions in the asymmetric region. In conclusion, when $N\gg1$, everywhere the maximal product overlap of W states is governed by the smallest $b_z$ among the $z$ components of the Bloch vectors. Using approximations
$$\frac{1}{2}\left(1+\s{1-\frac{c_i^2}{r^2}}\right)\approx e^{-c_i^2/4r^2},\quad i=1,2,\cdots,N-1$$
and equations \eqref{2.slight} and \eqref{4.r}  one obtains
\begin{equation}\label{5.g}
g^2(N\gg1)=
\begin{cases}
 \enskip \frac{1+b_z}{2}\,e^{-\frac{2b_z}{1+b_z}}, &{\rm if}\quad 0<b_z<\frac{1}{3}\cr
 \enskip \frac{1-b_z}{2}, & {\rm if}\quad b_z<0
\end{cases}
\end{equation}
 Graphic comparison of the interpolating formula and numerical computation of $g$ is shown in Fig.~\ref{int}, where the $b_z$ dependence of $g$ is plotted for $N=10$. The solid and dashed lines represent the interpolating function \eqref{5.g} and numerical computation, respectively.

We did not plotted numerical results for different states because different curves overlap and become indistinguishable. We failed to find the states for which the numerical results markedly differ from the plotted one provided $N\gg1$ holds.

\section{Discussion}

The main result of this work is the formula \eqref{5.g}. First, it shows that sometimes the characterization and manipulation of the entanglement of many qubit states is a simple task, while the case of few or several qubits is a complicated problem. Second, it states that when $N\gg1$ the maximal product overlap of W states is universal in the asymmetric and slightly entangled regions and the only exceptions are W states in the symmetric region that are almost maximally entangled states. Then a question arises: Why do the maximal product overlaps of the different W states far apart from the exceptional points have the same behavior?
Perhaps the reason is that these states are all W-class states.  Classification of entangled states explains that pure states can be probabilistically converted to one another within the same class by stochastic local operations and classical communication~\cite{w,four,bast}. And one can assume that large-scale systems within the same class have the feature, aside from the interconvertibility, that their entanglement is universal. An argument in favor of this assumption is that the geometric measure of entanglement~\cite{wei}, the relative entropy of entanglement~\cite{ved} and the logarithmic global robustness~\cite{robust} are related by bounding inequalities and, moreover, the relative entropy of entanglement is an upper bound to entanglement of distillation. Hence it is unlikely that these measures may exhibit contradicting results and each of them predicts its own and very different entanglement behavior of large-scale W-states. If this argument is true, then entanglement of large-scale states within the same class is universal. However, states from the different classes may exhibit different behaviors. By no means it is obvious, and probably not true, that the maximal product overlap of GHZ-class states should have a behavior similar to that of W states.

Another possible explanation is that the universality of the maximal overlap of large scale W states is the inherent feature of the geometric entanglement measure rather than the inherent feature of quantum states. If it is indeed the case, then a reasonable question is the following: do the exceptional points really exist or they are just the fabrication of the geometric entanglement measure? In this context the second exceptional point is a fundamental quantity. Indeed, there are states applicable for the perfect teleportation and dense coding and these states all should possess the same amount of entanglement. Hence there is an specific entanglement point(infinite entanglement diameter in the case of the geometric measure) that can be associated with the exceptional point. And one can assume that the second exceptional point is a property of quantum states rather than a property of the maximal product overlap. 	And how about the first exceptional point? Unfortunately, we do not know any strong arguments in favor of it. In order to clarify the existence or nonexistence of the first exceptional point, as well as the second exceptional point, one has to analyze another reliable entanglement measure, say relative entropy of entanglement~\cite{park}, and see whether it possesses exceptional points.

\begin{acknowledgments}
ST thanks Roman Orus Lacort for careful reading and feedback. This work was supported by ANSEF Grant No. PS-1852.
\end{acknowledgments}

\appendix
\section{Geometrical interpretation of the duality}
\begin{figure}[ht!]
\begin{center}
\includegraphics[height=7cm]{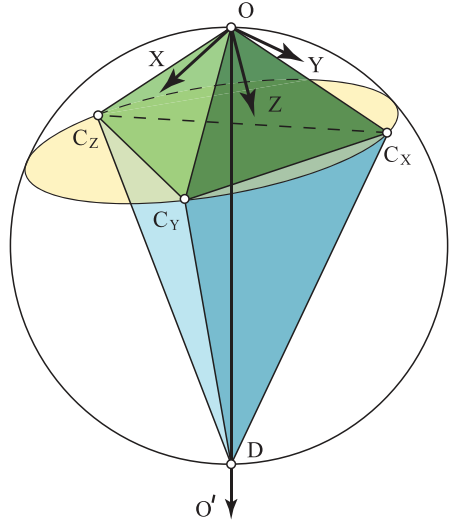}
\caption[fig7]{\label{vec}(Color online) The geometrical interpretation of the duality for three-qubit W states. Mutually perpendicular bold lines $OX$, $OY$ and $OZ$ are coordinate axes and  $\overrightarrow{OO^\prime}$ is an arbitrary direction. $OC_X,\,OC_Y$ and $OC_Z$ are mirror images of the line $OO^\prime$ in respect to the three axes. The points $C_X,\,C_Y$ and $C_Z$ are intersections of
these lines with the sphere uniquely defined by the two conditions: its center lies on the line $OO^\prime$  and its diameter $OD\equiv r$ is the sum of the lateral sides of the upper pyramid (with the apex $O$ and base $C_XC_YC_Z$). Now the direction cosines (and sines) of the vector $\overrightarrow{OD}$ are coefficients of the local states $\ket{u_i}$ in a computational basis. And the lateral sides of the lower pyramid (with the apex $D$ and base $C_XC_YC_Z$) are the coefficients of a 3-qubit W-state in the same basis. Thus each direction singles out a product state and a W state and thereby establishes a correspondence among them.}
\end{center}
\end{figure}
The nearest product state $\ket{u_1}\o\ket{u_2}\o\cdots\o\ket{u_N}$  of the W state \eqref{2.w} can be parameterized as follows
\begin{equation}\label{0.near}
 \ket{u_k}=\sin\t_k\ket{0}+\cos\t_k\ket{1},\, 0\leq\t_k\leq\frac{\pi}{2}, \, k=1,2,...,N,
\end{equation}
where
\begin{equation}\label{0.dircos}
\cos^2\t_1+\cos^2\t_2+\cdots+\cos^2\t_N=1.
\end{equation}
Thus the angles $\cos\t_k$ define a unit N-dimensional vector in Euclidean space. They satisfy the equalities \begin{equation}\label{0.rmod}
\frac{1}{r} \equiv \frac{\sin2\t_1}{c_1} = \frac{\sin2\t_2}{c_2} = \cdots =
\frac{\sin2\t_N}{c_N}.
\end{equation}
These equalities can be interpreted as trigonometric relations for the right triangles with hypotenuses $r$, angles $2\t_k$, opposite legs $c_k$ and adjacent legs $\s{r^2-c_k^2}$. If $2\t_k>\pi/2$, then one takes the angle $\pi-2\t_k$ instead. All of these triangles has the same hypotenuse $r$ and therefore can be circumscribed by a single sphere with the diameter $r$. The final picture represents two inscribed N-dimensional pyramids with a common base and lateral sides $c_1,c_2,...,c_N$ and $\s{r^2-c_1^2},\s{r^2-c_2^2},\s{r^2-c_N^2}$, respectively. The case $N=3$ is illustrated in Fig.\ref{vec}.

\end{document}